\documentclass[fleqn,a4paper]{article}
\usepackage{a4wide}
\usepackage{amsmath}
\usepackage{amssymb}
\usepackage{comment}
\usepackage{hyphenat}
\usepackage{multicol}
\usepackage{url}
\usepackage{mathrsfs}
\usepackage{graphicx}
\usepackage{microtype}                 
\PassOptionsToPackage{warn}{textcomp}  
\usepackage{textcomp}                  
\usepackage{mathptmx}                  
\usepackage{times}                     
\usepackage{cite}                      
\usepackage{tabu}                      
\usepackage{booktabs}                  

\newcommand {\mm}[1] {\ifmmode{#1}\else{\mbox{\(#1\)}}\fi}

\newcommand{\Xspace}        {\mm{{X}}}

\newcommand{\Ucal}{\mathcal{U}}
\newcommand{\Vcal}{\mathcal{V}}

\newcommand{\RV}               {\mathrm{RV}}
\newcommand{\WD}               {\mathrm{WD}}

\newcommand{\swissroll}	      {\emph{Swiss roll with a hole}}
\newcommand{\octa}               {\emph{Octa}}
\newcommand{\fishing}               {\emph{Fishing Net}}
\newcommand{\fourelt}               {\emph{4elt}}
\newcommand{\cylinderthree}     {\emph{Cylinder-3}}
\newcommand{\cylinderfive}        {\emph{Cylinder-5}}
\newcommand{\airfoil}		  {\emph{Airfoil1}}
\newcommand{\bcsstk}		  {\emph{Bcsstk31}}
\newcommand{\mice}		  {\emph{Mice}}

\newcommand{\qed}{\hfill$\Box$}

\title{Homology-Preserving Dimensionality Reduction \\via Manifold Landmarking and Tearing}
\author{Lin Yan\footnote{Lin Yan is with University of Utah. 
 E-mail: linyan@cs.utah.edu.}, Yaodong Zhao\footnote{ Yaodong Zhao is with University of Utah. E-mail: yaodong.zhao@utah.edu.}, Paul Rosen\footnote{Paul Rosen is with University of South Florida. E-mail: prosen@usf.edu.},  Carlos Scheidegger\footnote{Carlos Scheidegger is with University of Arizona. E-mail: cscheid@cs.arizona.edu. }, Bei Wang\footnote{Bei Wang is with University of Utah.   E-mail: beiwang@sci.utah.edu.}}

\providecommand{\keywords}[1]{\textbf{\textit{Index Terms---}} #1}
\begin{document}
\maketitle

\begin{figure}[tb]
 \centering 
 \includegraphics[width=10cm]{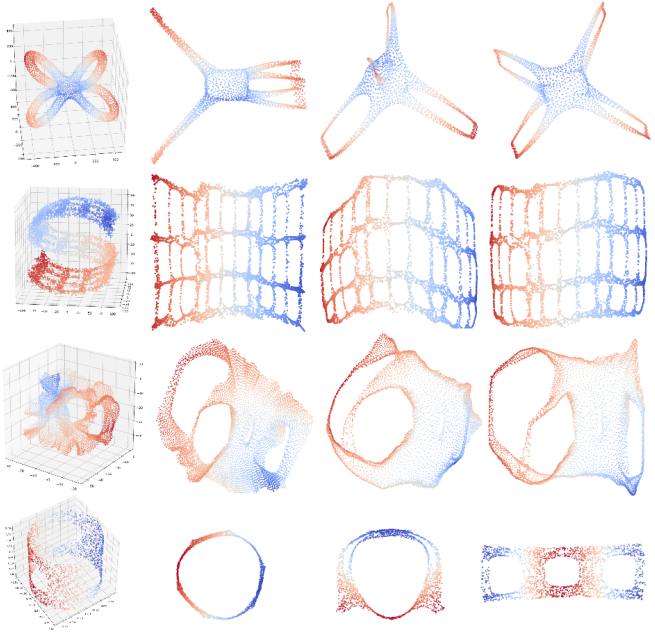}
 \caption{In search of homology-preserving linear projections for a \emph{bended figure eight} example. (a) Uniformly sample projection directions on a sphere. (b) The original 3-dimensional input point cloud. (c)-(e) Three instances of the 2-dimensional linear embeddings, (c) being the optimal as it produces the lowest degree-2 Wasserstein distortion.}
 \label{fig:teaser}
\end{figure}

\abstract{
Dimensionality reduction is an integral part of data visualization. It is a process that obtains a structure preserving low-dimensional representation of the high-dimensional data.  
Two common criteria can be used to achieve a dimensionality reduction: distance preservation and topology preservation. 
Inspired by recent work in topological data analysis, we are on the quest for a dimensionality reduction technique that achieves the criterion of homology preservation, a generalized version of topology preservation. 
Specifically, we are interested in using topology-inspired manifold landmarking and manifold tearing to aid such a process and evaluate their effectiveness. 
}


\section{Introduction}
\label{sec:introduction}
Dimensionality reduction (DR) is a process that obtains a \emph{structure-preserving} low-dimensional representation of the high-dimensional data.  
It plays an important role in high-dimensional data visualization in both static and interactive settings. 
Two common criteria can be used to achieve a DR \emph{distance preservation} and \emph{topology preservation}. 
Inspired by recent work in topological data analysis, we are on the quest for a DR technique that achieves the criterion of \emph{homology preservation}, a generalized version of topology preservation. 

To motivate our work, we begin by addressing the following questions: 
What is homology in the context of topology?  
What is homology preservation in the context of structure preservation?
Why does it matter (and who cares) for homology preservation in high-dimensional data analysis and visualization?
See the following paragraphs on a history of homology, a discussion on homology-preservation DR, and motivations from visualization to robotics.

\paragraph{A brief history of homology.}
Topology has been one of the most exciting research fields in modern mathematics~\cite{James1999}. 
It is concerned with the properties of space that are preserved under continuous deformations, such as stretching, crumpling, and bending, but not tearing or gluing~\cite{WikipediaTopology2018}. 

The beginning of topology is arguably marked by Leonhard Euler, who published a paper in 1736 that solves the now famous K\"{o}nigsberg bridge problem. 
In the paper, titled \emph{``The Solution of a Problem Relating to the Geometry of Position"}, Euler was dealing with ``a different type of geometry where distance was not relevant."~\cite{OConnorRobertson1996}  
Johann Benedict Listing was credited to be the first to use the word ``topology" in print based on his 1847 work titled \emph{``Introductory Studies in Topology"}; although many of Listing's topological ideas were due to Carl Friedrich Gauss~\cite{OConnorRobertson1996}.  
Both Listing and Bernhard Riemann studied the \emph{components} and \emph{connectivity} of surfaces. 
Listing examined connectivity in $3$-dimensional Euclidean space while Enrico Betti extended the idea to $n$ dimensions. 
Henri Poincar\'{e} then gave a rigorous basis to the idea of connectivity in a series of papers \emph{``Analysis situs"} in 1895. He introduced the concept of \emph{homology} and improved upon the precise definition of Betti numbers of a space~\cite{OConnorRobertson1996}. 
In other words, it was Poincar\'{e} who ``gave topology wings"~\cite{James1999} via the notion of homology. 

The original motivation to define homology was that it can be used to tell two things (a.k.a. topological spaces) apart by examining their holes. 
It is a process that associates a topological space with a sequence of abelian groups called homology groups, which, roughly speaking, count and collate \emph{holes} in a space~\cite{Ghrist2008}. 
In a nutshell, homology groups generalize a common-sense notion of connectivity. 
They detect and describe the connected components ($0$-dimensional holes), tunnels ($1$-dimensional holes), voids ($2$-dimensional holes), and holes of higher dimensions in the space. 

It is not easy to give homology an intuitive and correct definition, 
but we will attempt to give a layman's version by quoting and modifying one given by Evelyn Lamb~\cite{Lamb2014}: If it has one connected piece, it has a $0$-dimensional hole (imaging a cookie); If you can put it on a necklace, it has a $1$-dimensional hole; If you can fill it with toothpaste that is not exposed to the air (imaging a basketball), it has a $2$-dimensional hole; For holes of higher dimensions, you're on your own.

\paragraph{A discussion on homology-preservation.} 
DR techniques could be classified based on structure preservation, namely, distance preservation or topology preservation. 
The preservation of pairwise distances ensures that the low-dimensional embedding inherits the geometric properties of the data~\cite{GraciaGonzalezRobles2014}.  
For instance, classical Multidimensional Scaling (MDS, a linear technique) preserves spatial distances (such as the Euclidean distances) while Isomap (a nonlinear technique) uses geodesic distances (approximated by graph distances)~\cite{LeeVerleysen2007}. 
On the other hand, the quantitative natural of distance preservation also makes it very constraining -- it is like ``supporting and bolting the space with rigid steel beams"; and in nonlinear cases (such as manifolds), distances cannot be perfectly preserved~\cite{LeeVerleysen2007}. 
The notion of topology preservation refers to the preservation of neighborhood relations between subregions of the data. 
Topology preservation techniques, such as locally linear embedding (LLE)~\cite{RoweisSaul2000} and Laplacian eigenmap~\cite{BelkinNiyogi2003}, introduce some flexibility where subregions of the data could be locally stretched or shrunk in order to embed them in a lower dimensional space~\cite{LeeVerleysen2007}. 

In a way, many topology preservation DR techniques are concerned with a common-sense notion of connectivity; that is, the $0$-dimensional connectivities among neighboring data points. 
In this paper, we focus on a generalized version of topology preservation, \emph{homology preservation}, where we are interested in the preservation of both $0$-dimensional and $1$-dimensional homology of the data. 
In particular, we are in search of DR techniques that could preserve as much as possible of the $1$-dimensional homology (a.k.a.~\emph{loops}) of the data. 

\paragraph{Motivations from visualization to robotics.}
Our first motivation to study homology-preserving DR is from the perspective of visualization. 
As technologies advance, we are collecting and generating a wide variety of large, complex, and high-dimensional datasets that demand insight-generating analysis and visualization. 
However, limitations on our visual systems as well as display devices have prevented us from the rapid recognition of structures beyond three dimensions. Visualization approaches therefore play an essentially role in \emph{visually} conveying and interpreting high-dimensional structural information by utilizing low-dimensional embeddings and abstractions: from DR to visual encoding, and from quantitative analysis to interactive exploration~\cite{LiuMaljovecWang2017}. 
We believe homology-preserving DR helps to expand the existing DR toolset and encodes additional structural information of high-dimensional data for visual exploration.

Our second motivation is the availability of interesting datasets with nontrivial homology, in particular, from imaging and signal processing. 
In studying the space of images, Lee et al.~\cite{LeePedersenMumford2003} have found that the majority of high-contrast 3 by 3 patches are concentrated near a circle. 
Follow-up work by Carlsson et al.~\cite{CarlssonIshkhanovDe-Silva2008} 
and others~\cite{Xia2016} has shown that a subspace of the space of natural image patches either exhibits circular behavior or is  topologically equivalent to a Klein bottle, depending on the patch size. 
In signal processing, using delayed window embedding, a $1$-dimensional signal can be encoded into a high-dimensional point cloud  for topological data analysis. Specifically, $1$-dimensional homology (i.e.~loop) of such a point cloud captures the periodicity of the signal~\cite{PereaHarer2015, PereaDeckardHaase2015}. 

Finally, we are motivated by the large collection of datasets that arise from robotics. 
Homological concepts naturally arise in robotics from the perspective of \emph{motion planning}, that is, a process that aims to "design a  trajectory of robot states from a given initial state to a specified goal state through a complex configuration space"~\cite{HaParkChoi2016}. 
Two trajectories are considered being topologically equivalent, if the boundaries formed by both trajectories do not contain any obstacle. 
The notion of \emph{punctured Euclidean space}, that is, $R^D - O$, occurs frequently in the configuration spaces of robots, where $O$ represent either the physical obstacles (such as humans, chairs, or other robots) that the robots need to avoid in the $2$- or $3$-dimensional configuration spaces, or illegal states (such as all legs off the ground) for the higher-dimensional configuration spaces~\cite{BhattacharyaLipskyGhrist2013}.  
Homological information can be used to cluster and classify different classes of trajectories in complex configuration spaces and to find representative (optimal) trajectory for each class~\cite{BhattacharyaLipskyGhrist2013}. 
We envision homology-preserving DR techniques could help capture the nontrivial homologies in the environments for path planning and state planning. As part of future work, we are interested in knowing how topology-inspired landmarks and skeletons (Section~\ref{sec:landmarking}) can be used as representative trajectories in the configuration space. 

\paragraph{Contributions.}
The goal of this paper is to generalize topology preservation to homology preservation, much as generalizing connectivity to the notion of homology. 
Given that distance preservation maintains the geometric and therefore topological properties of the data with some sense of rigidity, this could be the end of the story. 

However, we present examples in the paper illustrating that we can achieve homology preservation while at the same time maintaining (and sometimes even improving) the preservation of distances. 
Our contributions are: 
\begin{itemize}
\item We introduce a new class of homology-preserving DR techniques that combine the strengths of landmark Isomap (L-Isomap) with the power of homology-preserving landmarks. 
\item For complex data such as circular manifolds, we provide a simple and fast procedure that can tear those manifolds, while at the same time preserving as much homology as possible. 
\item We conduct experiments for homology-preserving manifold landmarking and manifold tearing to evaluate their effectiveness. 
\end{itemize}

\paragraph{Outline.}
We motivate our problem of interest in Section~\ref{sec:linear-projection} with a naive interpretation of homology preservation in the setting of linear projection.  
We discuss related work in Section~\ref{sec:related-work} that covers topics such as manifold landmarking, manifold tearing, and quality assessment of DR techniques. 
We introduce homology-based measure for the assessment and comparison of DR techniques in Section~\ref{sec:evaluation}. 
We detail a novel landmark selection method for L-Isomap based upon data skeletonization, and describe a new class of homology-preserving DR techniques in Section~\ref{sec:landmarking}.
We describe homology-preserving manifold tearing in Section~\ref{sec:tearing}.
We conclude with real-world examples (Section~\ref{sec:results}) and a discussion on future directions (Section~\ref{sec:discussion}). 

\section{A Motivational Example Using Linear Projections}
\label{sec:linear-projection}

\begin{figure}[tb]
 \centering 
 \includegraphics[width=0.6\columnwidth]{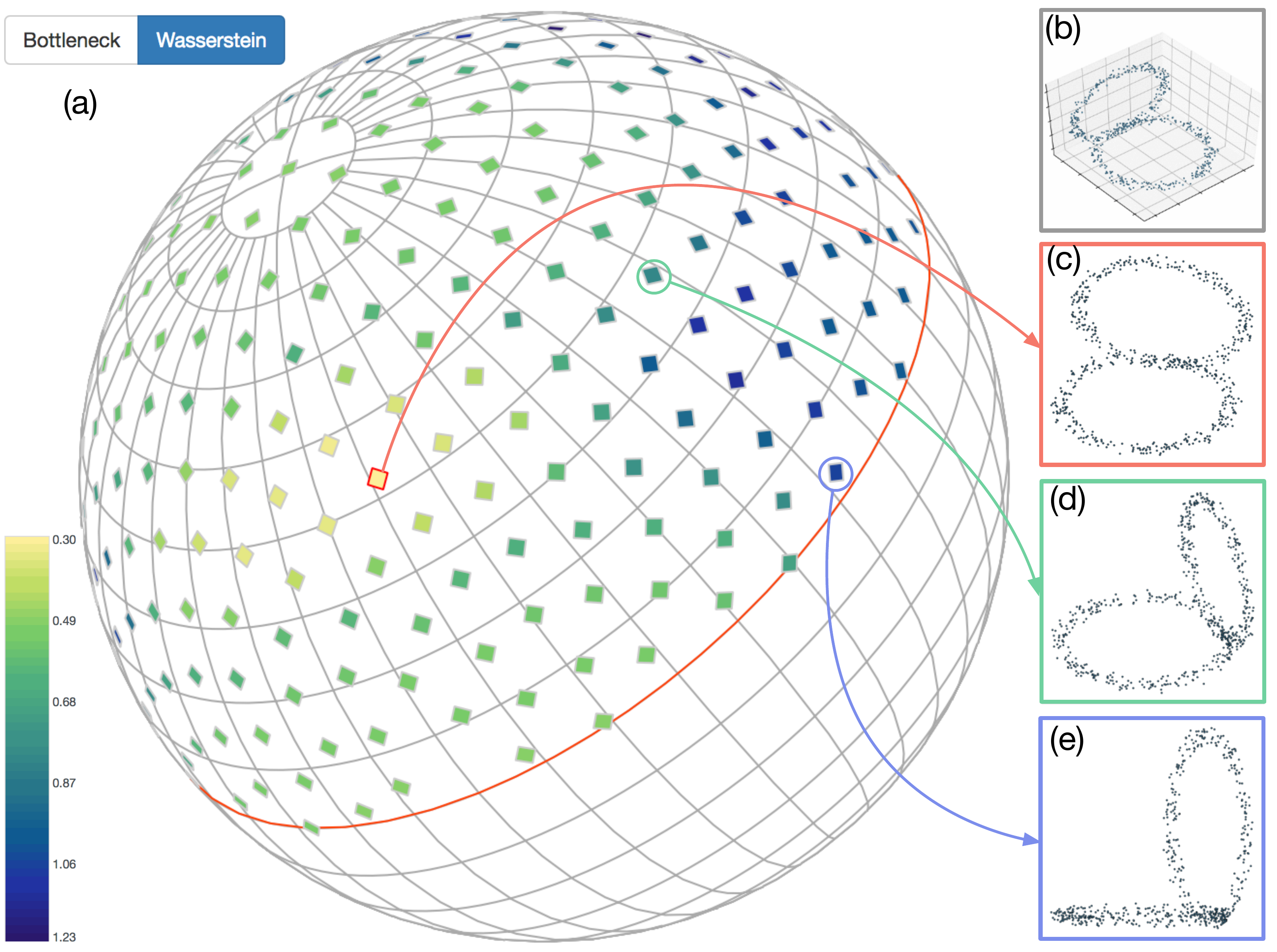}
 \caption{In search of homology-preserving linear projections for a \emph{bended figure eight} example. (a) Uniformly sample projection directions on a sphere. (b) The original 3-dimensional input point cloud. (c)-(e) Three instances of the 2-dimensional linear embeddings, (c) being the optimal as it produces the lowest degree-2 Wasserstein distortion.}
 \label{fig:linear-projection}
\end{figure}

To motivate our research objective, we start with a naive interpretation of homology-preservation DR using linear projects. 
Given a high-dimensional point cloud, we have created an interactive visualization  prototype, where we observe and quantify the results of DR using simple linear projections. 
We illustrate the idea of homology preservation using a $3$-dimensional point cloud sampled from two perpendicular rings joined at a point, referred to as the \emph{bended figure eight} in Figure~\ref{fig:linear-projection}(b). 
 
We begin by sampling a number of projection directions uniformly from a sphere in Figure~\ref{fig:linear-projection}(a). 
Given a particular projection direction, we use a couple of homology-centric criteria (detailed in Section~\ref{sec:evaluation}) to assess the quality of each linear embedding. 
Here, we focus on the preservation of $1$-dimensional homology. 
The optimal projection direction is marked by a red diamond in Figure~\ref{fig:linear-projection}(a), whereas Figure~\ref{fig:linear-projection}(c) corresponds to the optimal $2$-dimensional embedding. 
Figure~\ref{fig:linear-projection}(d) and (e) give two examples of not-so-optimal embeddings that correspond to the blue and green projection directions in (a), respectively. Each embedding shows a certain amount of 
deformation to at least one of the two loops. 

It is clear from the visualization that the optimal linear projection direction tries to preserve the shape of the two loops as much as possible, therefore  producing the most optimal homology preserving embedding. 

\section{Related Work}
\label{sec:related-work}

\paragraph{Dimensionality reduction.}
DR techniques can be studied following various taxonomies. 
For instance, they are considered as linear (resp. nonlinear) methods if they produce low-dimensional linear (resp. nonlinear) mapping of the input high-dimensional data that preserve certain features of interest. 
They can be thought of as conducting convex or nonconvex optimizations, full or sparse spectral eigendecompositions, global or local structure preservation; see~\cite{CunninghamGhahramani2015, MaatenPostmaHerik2007} for thorough reviews. 
We largely follow the classification from~\cite{LeeVerleysen2007} in terms of distance or topology preservation (see Section~\ref{sec:introduction}). 

\paragraph{Quality assessment and visualization.}
To assess the performance of DR techniques, different quality measures have been proposed that can be roughly classified as global- or local-based approaches. The former quantifies the preservation of local neighborhoods/subregions,  and the latter studies the preservation of global shape of data. 
Global measures include Shepard diagram~\cite{Shepard1962, Shepard1962b}, stress~\cite{Kruskal1964, Kruskal1964b}, and residual variance~\cite{TenenbaumSilvaLangford2000} (as described in Section~\ref{sec:evaluation}), and local measures consist of rank-based criteria such as co-ranking matrix~\cite{LeeVerleysen2009}, mean relative rank errors~\cite{LeeVerleysen2009} and Spearman's rho~\cite{SiegelCastellan1988}, 
normalization independent embedding quality assessment~\cite{ZhangRenBoZhang2012}, and many more~\cite{GraciaGonzalezRobles2014}. 

On the other hand, the characteristics of various quality measures can be  linked with fine-grained visual analytics. 
For instance, point-wise quality measures can be augmented in the visualization to highlight erroneous local regions~\cite{MokbelLueksGisbrecht2013}. 

\paragraph{Manifold landmarking.}
Our proposed strategy takes advantage of \emph{manifold landmarking}, that is, finding a subset of points along the manifold that captures its structural characteristics~\cite{LiangPaisley2015}.  
Manifold landmarking is useful for dimensionality reduction, for example, in the case of landmark MDS and landmark Isomap~\cite{SilvaTenenbaum2003, SilvaTenenbaum2004}. 
It can also be employed to generate sparse manifolds for machine learning tasks~\cite{NascimentoCarneiro2017} or sparse matrices for semidefinite programming~\cite{WeinbergerPackerSaul2005}, as well as supervised learning~\cite{Tipping2001}. 

Previous landmarking methods can be classified as geometric or statistical approaches. 
Geometrically, landmarks can be selected randomly~\cite{SilvaTenenbaum2003} or using \emph{maxmin} methods~\cite{SilvaTenenbaum2004}. 
Selection can focus on the boundaries using minimum spanning tree, rather than randomly selected points or cluster centers~\cite{ChenCrawfordGhosh2006}. 
Landmarks can also be chosen based on the minimum set cover problem~\cite{LeiYouDong2013, ShiYinBao2016}. 
In addition, modification to existing landmark-based DR method such as L-Isomap can be done by introducing modifications to the distance matrices involving landmarks and its related spectral problems~\cite{ShiHeLiu2015, VladymyrovCarreira-Perpinan2013}.  
Statistically, manifold landmarking utilizes regression~\cite{SilvaMarquesLemos2006}, 
mixed-integer optimization~\cite{OrsenigoVercellis2013}, active learning~\cite{XuYuDavenport2017}, and Gaussian processes~\cite{LiangPaisley2015}. 
  
\paragraph{Topology-Inspired data skeletonization.}  
Compared to exiting landmarking approaches, our strategy is one that is topological in nature. 
Our work utilizes advances in topology-inspired data skeletonization, that is, the process of extracting the topological structure of data using a low-dimensional even $1$-dimensional representation,  
in order to better interpret complex, noisy, nonlinear, and high-dimensional data. 

Data skeletonization can be broadly considered as a graph-extraction problem.
The work in~\cite{HastieStuetzle1989, KeglKrzyzak2002} develops the notions of \emph{principal curves} and \emph{principal graphs}, which are roughly smooth curves that pass through the middle of a cloud of points. 
Metric graph reconstruction~\cite{AanjaneyaChazalChen2012} also helps to skeletonize data based on inspecting and classifying local neighborhoods.  

Topology-inspired data skeletonization from~\cite{GeSafaBelkin2011, NataliBiasottiPatane2011} are most relevant to our framework. 
Ge et al.~\cite{GeSafaBelkin2011} give a framework to extract and simplify a $1$-dimensional skeleton using the Reeb graph.
Natali et al.~\cite{NataliBiasottiPatane2011} introduce a \emph{Point Cloud Graph} as a data abstraction that is a generalization of the Reeb graph to arbitrary high-dimensional point clouds.  
Reeb graphs play a fundamental role in computational topological, topological data analysis and shape analysis; see~\cite{BiasottiGiorgiSpagnuolo2008} for a survey. 

In this paper, we extract a $1$-dimensional skeleton (referred to as \emph{skeleton} for the remaining of the paper)  from the input space based on an approximation of the Reeb graph. Compared to previous work, our work is novel in the sense that it utilizes such a skeleton for the purpose of landmark section and DR. 

\paragraph{Manifold tearing and loop detection.}
Most classic DR techniques do not perform well when the data manifolds contain essential (i.e.,~non-contractable) loops, such as cylinders, tori or spheres. 
What sometimes are referred to as loopy manifolds~\cite{MengLeungXu2013} are in fact manifolds with nontrivial homology. 
Such manifolds typically cannot be embedded into the target space without introducing significant distortions. 
Some recent efforts have been made to detect and cut essential loops in such manifolds. 

Lee and Verleysen~\cite{LeeVerleysen2005} introduce a two-stage tearing procedure: first, a k-nearest neighbor (kNN) graph among the point cloud sample is used to represent the underlying space; second, a minimum spanning tree (MST) or a shortest path tree (SPT) that contains no cycles is computed on the kNN graph; Finally, edges that do not generate non-contractible cycles with more than 4 edges are reintroduced to form the torn graph for downstream DR. 

Our work differs from~\cite{LeeVerleysen2005} significantly in the following sense. We use a topology-inspired data skeleton that consists of landmarks and landmark connections to describe all candidate essential loops, and employ a homological criterion to choose the proper loop to tear while preserving as much as possible the homological characteristics of the data. 
Whereas other techniques cut all or a large number of loops, 
we try to cut, roughly, as few loops as possible while preserving the remaining homology.


\section{Homology-Based Quality Assessment}
\label{sec:evaluation}

\subsection{Background}
Before we present various homology-based criteria in assessing the quality of DR, we need to introduce a few relevant topological notions such as homology and persistent homology~\cite{EdelsbrunnerHarer2008}, and a distance-based evaluation criterion.

\paragraph{Homology and Betti numbers.}
Given a topological space X, the $0$-, $1$- and $2$-dimensional homology groups are denoted as $H_0(X)$, $H_1(X)$ and $H_2(X)$, respectively. 
Betti numbers $b_i$ count the number of $i$-dimensional holes, and are used to distinguish spaces based on the connectivity across all dimensions. 
Formally, it is defined as the rank of the $i$-dimensional homology groups, $b_i = rank(H_i(X))$.  
For a torus, $b_0 = 1$,  $b_1 = 2$ and $b_2 = 1$; this means that a torus has $1$ connected components, $2$ holes and $1$ void. 

\begin{figure}[ht!]
\centering 
\includegraphics[width=0.6\columnwidth]{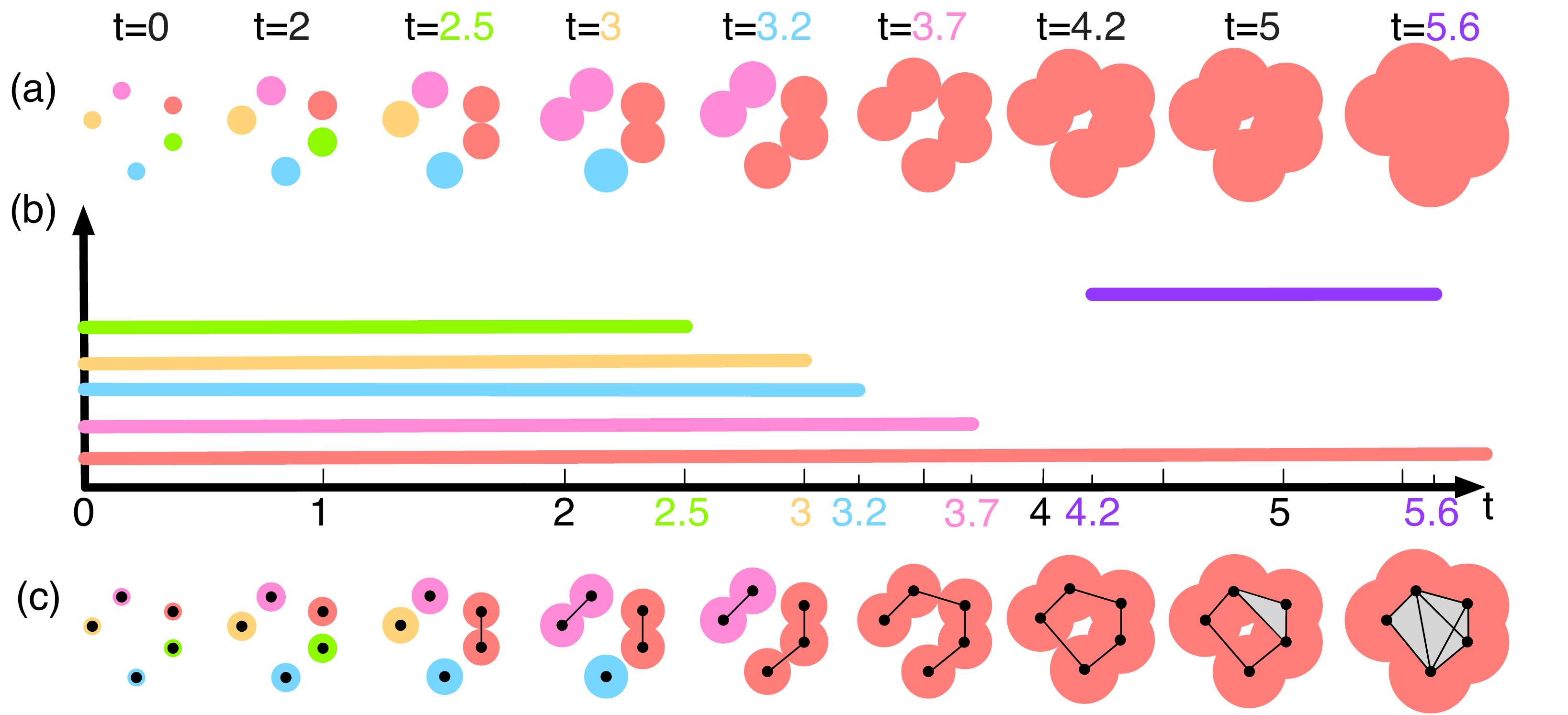}
\vspace{-4mm}
\caption{Computing persistent homology of a point cloud and the barcode.}
\label{fig:persistent-homology}
\end{figure}

\paragraph{Persistent homology.}
Simply put, persistent homology studies homology at multiple scales. 
As illustrated in Fig.~\ref{fig:persistent-homology}, we begin with a point cloud $X$ equipped with a distance metric $D_X$ (i.e.~Euclidean distance). 
We study the homology of a sequence of spaces formed by a union of balls of increasing radius $t$ centered at the points.  
Using persistent homology, we investigate the homological changes within this  growing sequence of spaces indexed by time (this is referred to as a \emph{filtration}).

\begin{figure}[ht!]
\centering 
\vspace{-4mm}
\includegraphics[width=0.5\columnwidth]{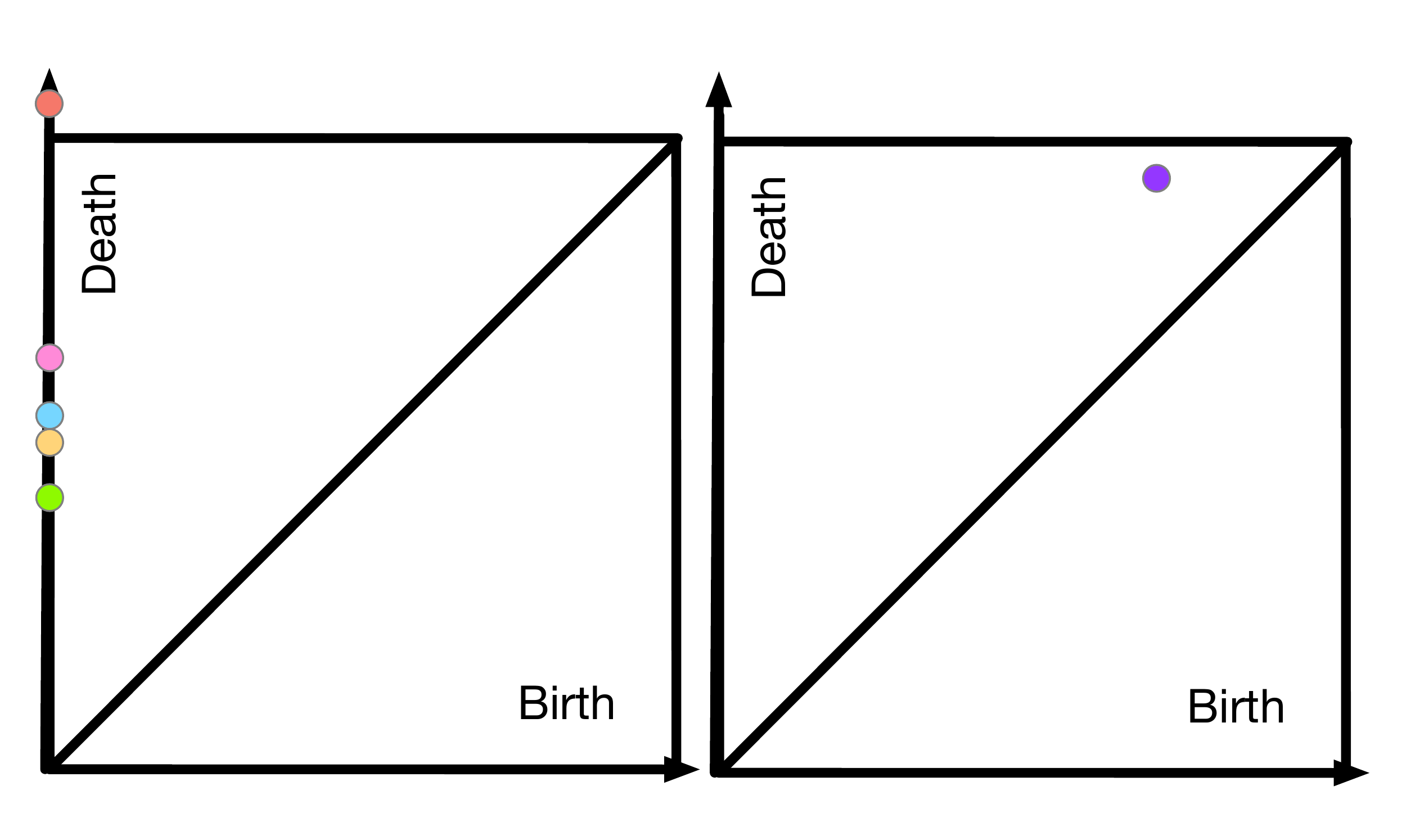}
\vspace{-6mm}
\caption{From left to right, $0$- and $1$-dimensional persistence diagrams.}
\label{fig:pd}
\end{figure}

In Fig.~\ref{fig:persistent-homology}(a), at time $t = 0$, each colored point is \emph{born} (appears) as its own (connected) component. As $t$ increases, we focus on the important events when components merge with one another to form larger components or tunnels. 
We begin by tracking the birth and death times of each component or tunnel as well as its lifetime in the filtration.
At $t = 2.5$, the green component merges into the red component and \emph{dies} (disappears); therefore the green component has a lifetime (i.e., \emph{persistence}) of $2.5$.
At $t = 3$, the orange component merges into the pink component and dies; therefore it has a persistence of $3$.
Similarly, the blue component dies at $t = 3.2$ while the pink component dies at $t = 3.7$.
At time $t = 4.2$, the collection of components forms a tunnel; and the tunnel disappears at $t = 5.6$. 
The red component born at time $0$ never dies, therefore it has a persistence of $\infty$.
We record and visualize the appearance (birth), the disappearance (death), and the persistence of homological features in the filtration via persistence diagrams~\cite{Cohen-SteinerEdelsbrunnerHarer2007} (Fig.~\ref{fig:pd}), or equivalently, persistence \emph{barcodes}~\cite{Ghrist2008b}. 
A point $p=(a,b)$ in the persistent diagram of $X$ records a homological feature that is born at time $a$ and dies at time $b$.
$0$- and $1$-dimensional persistence diagrams, denoted as $PD_0(D_X)$ and $PD_1(D_X)$, captures the births and deaths of components and tunnels, respectively.  
Equivalently in the barcode of Fig.~\ref{fig:persistent-homology}(b), such a feature is summarized by a horizontal bar that begins at $a$ and ends at $b$.

Computationally, the above nested sequence of spaces can be combinatorially represented by a nested sequence of simplicial complexes (i.e.~collections of vertices, edges and triangles) with a much smaller footprint, as illustrated in Fig.~\ref{fig:persistent-homology}(c), see~\cite{EdelsbrunnerHarer2010} for computational details.

\paragraph{Distance-based quality measure.}
To evaluate the fits of various DR techniques on comparable grounds, 
Tenenbaum~et al.~\cite{TenenbaumSilvaLangford2000} introduce the \emph{residual variance} ($RV$): 
$$RV(X, Y) = 1 - R^2(D_X, D_Y).$$
$D_Y$ is the matrix of Euclidean distances in the low-dimensional embedding produced by a DR technique, and $D_X$ is a best estimate of the intrinsic manifold distance for a given technique. 
In the case of Isomap, $D_X$ corresponds to the geodesic distance matrix approximated by the graph distance matrix $D_{G}$. 
For MDS, it is the Euclidean distance from the input data. 
$R$ is the standard Pearson correlation coefficient that measures the linear correlation between all entries of $D_X$ and $D_Y$ (that are reshaped into vectors). 
Such a measure can also be used as a practical approach to select an appropriate neighborhood size for noiseless and noisy data, where lower residuals indicate betterfitting solutions with less metric distortion~\cite{BalasubramanianSchwartzTenenbaum2002}. 

\subsection{Homology-Based Quality Measures}

There have been a few recent works in the direction of a homology-based evaluation criterion for DR.   
Paul and Chalup~\cite{PaulChalup2017} compare the Betti numbers ($b_0$ and $b_1$) of spaces before and after DR and study their convergence as the number of sample points increases. 
Rieck and Leitte~\cite{RieckLeitte2015} present an evaluation scheme based on persistent homology. 
Specifically, density functions of the input space and the embedding space are estimated based on neighborhood graphs in both spaces; then persistence diagrams are computed for these density functions. 
The global quality of each embedding is assessed by computing the 
degree-$2$ Wasserstein distance between the two persistence diagrams. 
The local quality at each embedded point is estimated based on a mean matching cost between the diagrams. 

For the purpose of this paper, we employ a few homology-based quality measures described below.   

\paragraph{Wasserstein distance.}
Our first criterion is the Wasserstein distance for persistent homology~\cite{Cohen-SteinerEdelsbrunnerHarer2010}. 
Recall from the previous section, we could obtain a $i$-dimensional persistence diagram $PD_i(D_X)$ from a point cloud $X$ described by a pair-wise distance matrix $D_X$.  
Let $D_X$ and $D_Y$ denote the distance matrices describing the structure relations among points within the high-dimensional input space and those within the low-dimensional embedding, respectively. 

To measure similarities between the persistent homology of space $X$ and $Y$, for a fixed dimension $i$, we define the \emph{degree-$p$ Wasserstein distance} as the Wasserstein distance between the $i$-th persistence diagrams of $D_X$ and $D_Y$: 

\begin{center}
$W_p(X,Y) = \displaystyle \left[ \inf_{\eta}  \Sigma_{x} \left\lVert x-\eta(x) \right\rVert^p_\infty \right]^{1/p}$, 
\end{center}

where the infimum is over all bijections $\eta: PD_i(D_X)\rightarrow PD_i(D_Y)$, and the sum is over all points 
$x \in PD_i(D_X)$~\cite{Wasserstein1969}. 
Note that taking the limit $p \to \infty$ yields the bottleneck distance~\cite{Reininghaus2015}. 

In the context of this paper, we always use $p=2$, leading to the degree-$2$ \emph{Wasserstein distance} measure, denoted as $WD$. 
Depending on the chosen dimension $i$, $WD_0$ and $WD_1$ roughly capture the homological distortions of $0$- and $1$-dimensional features before and after DR.  
For the assessment of Isomap and L-Isomap, $D_X$ is the Euclidean distance matrix in the input space and $D_Y$ is the Euclidean distance matrix in the embedding. 

Similarly, we could consider the \emph{bottleneck distance} by taking the limit $p \to \infty$, 
$
W_{\infty}(X,Y) =  \inf_{\eta}  \Sigma_{x} \left\lVert x-\eta(x) \right\rVert_\infty. 
$
As illustrated in Figure~\ref{fig:linear-projection}, we find the (near) optimal linear embedding by minimizing the Wasserstein distance measure (resp. bottleneck distance measure). For the remaining of this paper, we use the Wasserstein distance. 

\paragraph{Persistent Betti numbers.}  
Betti numbers count the number of homological features and can be used as summary statistics to differentiate topological spaces. 
However, Betti numbers alone do not differentiate between significant and noisy homological features. 
We use the \emph{persistent Betti numbers} ($PB$) as a way to quantify how much homological information is preserved during the DR.
Let $PB_i$ denote the number of significant $i$-dimensional homological features, that is, the number of points in the $i$-dimensional persistence diagram that is above a certain threshold that separates features from noise. 

Finding a suitable threshold requires checking the separation of points in the persistence diagram~\cite{RieckLeitte2015b}. 
Significant features can be extracted from the persistence diagram if it has an empty band (of a certain width) parallel to the diagonal that does not contain any points~\cite{ChazalGuibasOudot2013}. 
More sophisticated methods from statistics based on bootstrapping can be used to improve the threshold estimation that separate signals from noise, based on the notion of a \emph{confidence band}~\cite{FasyLecciRinaldo2014}, this is left to the future work. 

In this paper, we use $PB_1$ as a rule of thumb to assess the quality of DR in terms of its preservation of significant $1$-dimensional features.  

\paragraph{Comparisons to prior work.} 
We compare our DR framework with the prior work that use homology-based quality measures~\cite{PaulChalup2017, RieckLeitte2015}. 
Although it is generally true that data with ``more holes require higher sample sizes"~\cite{PaulChalup2017}, we demonstrate (in Section~\ref{sec:results}) that, by using carefully selected landmarks, we can preserve holes even with a small number of landmark points. 

Comparatively speaking, there are a few critical differences between our work and the work in~\cite{RieckLeitte2015}. 
The global and local quality measures from~\cite{RieckLeitte2015} deal only with $0$-dimensional while our proposed measure focuses on $1$-dimensional homology preservation. 
Most importantly, \cite{RieckLeitte2015} uses homology-based quality measures for the evaluation of existing DR techniques, while our work is actively searching and developing new DR techniques that maximize homology-based quality measures via manifold landmarking and tearing. 

\section{Homology-Preserving Manifold Landmarking}
\label{sec:landmarking}
We begin this section with a review of the nonlinear DR techniques known as the Isomap~\cite{TenenbaumSilvaLangford2000} and landmark Isomap (L-Isomap)~\cite{SilvaTenenbaum2003}. 
We then discuss homology-preserving landmark selection based on the Reeb graph and its discrete approximation. 
We summarize the pipeline for a new class of techniques by combining the utilities of homology-preserving landmarks with the efficiency of landmark-based DR. 

\subsection{Isomap and L-Isomap}

\paragraph{Isomap.} 
Suppose the original input data contains $N$ samples in $D$ dimensions, $X \in R^{D \times N}$. 
Isomap embeds the points onto a lower dimensional space $Y\in R^{d \times N}$ ($d < D$) while preserving geodesic distances between all input points~\cite{TenenbaumSilvaLangford2000}:
\begin{itemize}
\item[1.] \emph{Construct neighborhood graph}. A weighted, undirected $k$-nearest neighbor (kNN) graph $G$ is constructed over all data points, where an edge between a point $x_i \in X$ and its neighbor $x_j \in X$ is assigned a weight that represents the Euclidean distance between them. 
An appropriate $k$ can be chosen based on the residual variance~\cite{TenenbaumSilvaLangford2000, BalasubramanianSchwartzTenenbaum2002}.  
\item[2.] \emph{Compute shortest paths}. All pairwise shortest paths between points in the KNN graph $G$ are computed to approximate the geodesic distances between them, which leads to an $N \times N$ graph distance matrix $D_{G}$.  
\item[3.] \emph{Construct a $d$-dimensional embedding}. Classical MDS is applied to the above graph distance matrix $D_{G}$ to obtain a low-dimensional embedding. 
\end{itemize}
Isomap suffers from two computational inefficiencies: calculating the shortest-paths distance matrix and eigenvalues within MDS. 
The former has a complexity of $O(kN^2\log{N})$ using Dijkstra's algorithm with Fibonacci heaps, while the latter takes $O(N^3)$~\cite{SilvaTenenbaum2003}.

\paragraph{L-Isomap.} 
L-Isomap~\cite{SilvaTenenbaum2003} addresses the two inefficiencies of Isomap at once. It is based on the landmark MDS (L-MDS)~\cite{SilvaTenenbaum2004}: 
\begin{itemize}
\item [1.] \emph{Construct neighborhood graph} (same as in Isomap). 
\item [2.] \emph{Select landmarks}. $n$-points ($n \ll N$) from $X$ are randomly selected to be landmark points.  
\item [3.] \emph{Compute shortest paths.} Compute the shortest paths from each data point to the landmarks, resulting in a $n \times N$ geodesic distance matrix. Also compute the $n \times n$ shortest paths distance matrix between pairs of landmarks. 
\item [4.] \emph{Apply L-MDS to obtain $d$-dimensional embedding}. First, apply classical MDS to the landmarks only, embedding them in $R^d$ using as input the $n \times n$ distance matrix between pairs of landmarks. Second, the embedding coordinates for the remaining data points are computed based on a fixed linear transformation of their geodesic distances to the landmarks~\cite{ShiYinKang2017}. 
\item [5.] \emph{PCA normalization (optional)}. This normalization is to re-orient the axes of the embedding to reflect the overall distribution, rather than the distribution of the set of landmarks; see~\cite{SilvaTenenbaum2003, SilvaTenenbaum2004} for details.
\end{itemize}
L-Isomap leads to enormous savings when $n \ll N$:  
Computing the shortest paths in step 3 takes $O(knN\log{N})$ using Dijkstra's algorithm and L-MDS in step 4 runs in $O(n^2N)$~\cite{SilvaTenenbaum2003}. 

Here, to differentiate different versions of L-Isomap based on various landmark selection schemes, L-Isomap using randomly selected landmarks is referred to as the \emph{random L-Isomap}, while the one using homology-preserving landmarks in the next section is called the \emph{homology L-Isomap}. 
 
\subsection{Homology-Preserving Landmark Selection}
Our work uses the idea of a data skeleton based on the Reeb graph for the purpose of landmark selection in DR. 
Although Reeb graphs have been used in the context of shape abstraction and comparison~\cite{GeSafaBelkin2011, NataliBiasottiPatane2011}, to the best of our knowledge this is the first time they have been used in the context of landmark-based DR. 

In this section, we first review relevant topological notions and computations for Reeb graphs. We then describe our landmark section algorithm using a skeleton based on the Reeb graph. 

\paragraph{Reeb graph.}
Let $f: X \to R$ be a continuous function defined on a manifold $X$.  The level set of $f$ at a value $a \in R$ is defined as $f^{-1}(a) = \{x \in X \mid f(x) = a\}$. 
The \emph{Reeb graph} of $f$ is constructed by identifying every connected component in a level set to a single point~\cite{GeSafaBelkin2011}. 

\paragraph{Extracting homological skeleton}
Given point cloud data, the domain can be approximated by a neighborhood graph (such as the kNN graph or the $\epsilon$-neighborhood graph) among the data points, and efficient algorithms exist~\cite{HarveyWangWenger2010, GeSafaBelkin2011, NataliBiasottiPatane2011} to approximate the Reeb graph in such a discrete setting.  

In this paper, we employ a mapper-based implementation to approximate the Reeb graph~\cite{SaulVeen2017} as our homology-preserving data skeleton, refered to as the \emph{homologicla skeleton} (or simply skeleton). 
The mapper algorithm~\cite{SinghMemoliCarlsson2007} approximates the Reeb graph by considering the connected components of interval regions (i.e.~$f^{-1}(a,b)$) instead of the connected components of level sets (i.e.,~$f^{-1}(a)$). 

We start with a function $f: X \to R$ defined on a point cloud $X$, and a cover $\Ucal$ of $f(X)$ consisting of finitely many open intervals $\Ucal = \{(a_i, b_i)\}_{i=1}^{n}$. 
To specify such as cover, we pick two resolution parameters $n$ and $p$, where $n$ is the number of intervals and $p$ is the percentage of overlap between a pair of adjacent intervals. 
\emph{Pulling back} the cover $\Ucal$ through $f$ gives an open cover of the point cloud $X$, which is then refined into a connected cover by splitting each cover element into various clusters using a user-defined clustering algorithm~\cite{CarriereMichelOudot2017}. 
Such a cover of $X$ is denoted as $\Vcal = f^*(\Ucal)$. 
In this paper, we use DBSCAN~\cite{EsterKriegelSander1996} for clustering. It is a widely used density-based clustering algorithm that groups together points that are closely packed together; the choice of clustering algorithm is not essential to our experiments. 

The $1$-dimensional skeleton of the nerve of $\Vcal$ is considered a discrete approximation of the Reeb graph of $f$ on $X$; it is referred to as the \emph{homological skeleton} for the remainder of the paper. 
Such a skeleton is a graph with nodes representing the elements of $\Vcal$, and edges representing the pairs of cover elements in $\Vcal$ with nonempty intersections.

In the original mapper algorithm, the node of a skeleton represents abstractly a cover element of the point cloud, that is, a cluster of points in $X$. 
However, in our setting, the nodes of a homological skeleton are considered as the landmarks for DR; therefore, we choose the \emph{centroid} of each cluster as its representative. 
The centroid of each cluster is part of the original point cloud, and serves as a landmark for L-Isomap.  

Extracting homological skeleton does not increase the asymptotic complexity of L-Isomap. 
In our experiments, the running time of H-L-Isomap is comparable with that of R-L-Isomap. 

\paragraph{Filter function.}
The key idea behind the Reeb graph is that it explores the topology of a space by analyzing the behavior of a possibly varying real-valued function (referred to as a \emph{filter function}) defined on it~\cite{BiasottiGiorgiSpagnuolo2008}. 
Reeb graphs encode topological information on data in a $1$-dimensional structure, disregarding the dimension of the data in the ambient space~\cite{BiasottiGiorgiSpagnuolo2008}. 
The data can be regarded as being parameterized with respect to the filter function being used, in other words, the filter function ``plays the role of the \emph{lens}" through which we look at the properties of the data~\cite{BiasottiGiorgiSpagnuolo2008}. 

Different filter functions lead to different insights into the point cloud. 
It remains an open question as how to choose an appropriate filter function beyond a best practice or a guesstimation. 
Commonly used options include height functions, distances from the barycentre of a space, surface curvature, integral or average  geodesic distances and geodesic distances from a source point in the space~\cite{BiasottiGiorgiSpagnuolo2008}. 

In this paper, we use mainly the geodesic distance from a source point as the filter function, referred to as the distance-to-the-base-point or simply the $DTB$ function. 
Such a filter function has shown desirable properties in capturing the $1$-dimensional homological information of the space~\cite{BiasottiGiorgiSpagnuolo2008,GeSafaBelkin2011}. 
We demonstrate via experiments in Section~\ref{sec:results} that a skeleton induced by $DTB$ is homology-preserving for landmark-based DR; it also serves as a compact and informative summary for guiding the manifold tearing process. 

\paragraph{Landmark selection pipeline.} 
Given a point cloud $X$ in $R^D$, 
we now summarize our homology-preserving landmark selection pipeline and its combination with L-Isomap (referred to as the homology L-Isomap): 
\begin{itemize} 
\item[1.] \emph{Construct neighborhood graph} (same as Isomap).  Let $G$ denote the resulting kNN graph.
\item[2.] \emph{Compute a filter function $f$ on $X$}. $f$ captures certain desirable structural information of $X$ suitable for DR. In our experiments, we use $DTB$ as the filter function and a base point is chosen from extreme points or barycenters (see Section~\ref{sec:results} for details). $DTB$ can be computed based on $G$ from a given base point. 
\item[3.] \emph{Compute skeleton and landmarks.} 
Compute a discrete approximation of the Reeb graph of $f$ as a homological skeleton, using the mapper algorithm. 
The cover $\Ucal$ of $f(X)$ is given by user-specified  resolution parameters $n$ and $p$.  
The nodes of the skeleton correspond to clusters of points in $X$; the cluster centroids are chosen as the landmarks for L-Isomap, denoted as $X_L \subseteq X$. 
\item[4.] \emph{Apply L-Isomap}. Replace randomly generated landmarks (step 2 of L-Isomap) with the homology-preserving landmarks (a.k.a.~homological landmarks) $X_L$ and apply the rest of L-Isomap algorithm. 
\end{itemize}
The above pipeline is not restricted to L-Isomap.  We believe it can be easily extended to other graph-based DR techniques~\cite{YanXuZhang2007}. 

\begin{figure}[tb]
 \centering 
 \includegraphics[width=0.8\columnwidth]{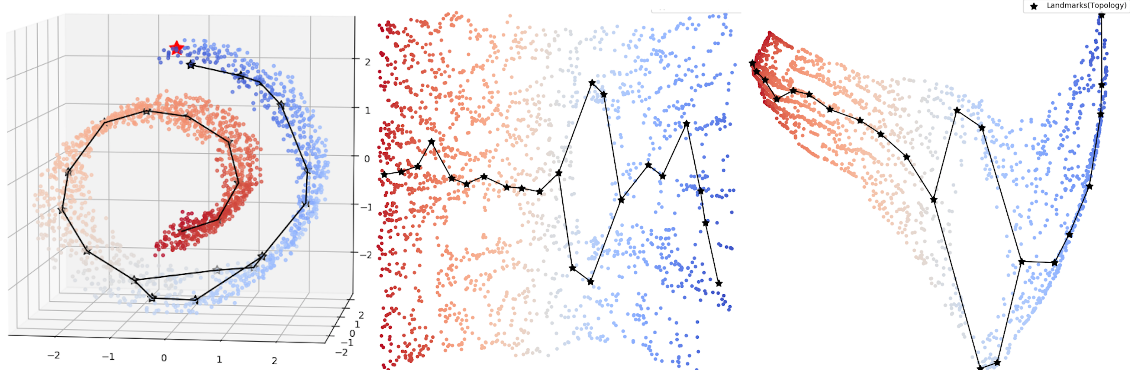}
 \caption{A \emph{swiss roll with a hole} example. (a) The original point cloud is colored by the \emph{distance to a base point} function. The base point is marked by a red star. (b) Isomap embedding. (c) Homological L-Isomap embedding. The homological skeleton is highlighted in black.}
 \label{fig:swiss-roll-hole}
\end{figure}

\paragraph{A simple example.}
Our pipeline is illustrated with a simple example in Fig.~\ref{fig:swiss-roll-hole}. 
We begin with a noisy point cloud with $1983$ points sampled from a swiss roll with an irregular, hard-to-spot hole in the middle.
First, a $DTB$ filter function is computed with respect to an extremal base point. Second, a homological skeleton connecting a set of $22$ landmarks is obtained using the mapper approximation. 
Third, we apply L-Isomap using these homological landmarks in the skeleton (black stars). 
In Fig.~\ref{fig:swiss-roll-hole}, the black homological skeleton is highlighted in the input space, as well as the Isomap and L-Isomap embeddings, which clearly captures the location of the significant hole in the data. 

This simple example demonstrates that the homological L-Isomap has the potential to preserve as much as possible the $1$-dimensional homological feature even with a smaller number of landmarks than the Random L-Isomap algorithm (roughly $n = O(\sqrt{N})$). 
However, given the dataset contains a single loop, homological L-Isomap does introduce distance distortion away from the loop. 
Surprisingly, homological L-Isomap is shown to outperform both Isomap and random L-Isomap in certain datasets using the widely accepted residual variance (see Section~\ref{sec:results}). 

\begin{figure*}[tb]
 \centering 
 \includegraphics[width=0.8\textwidth]{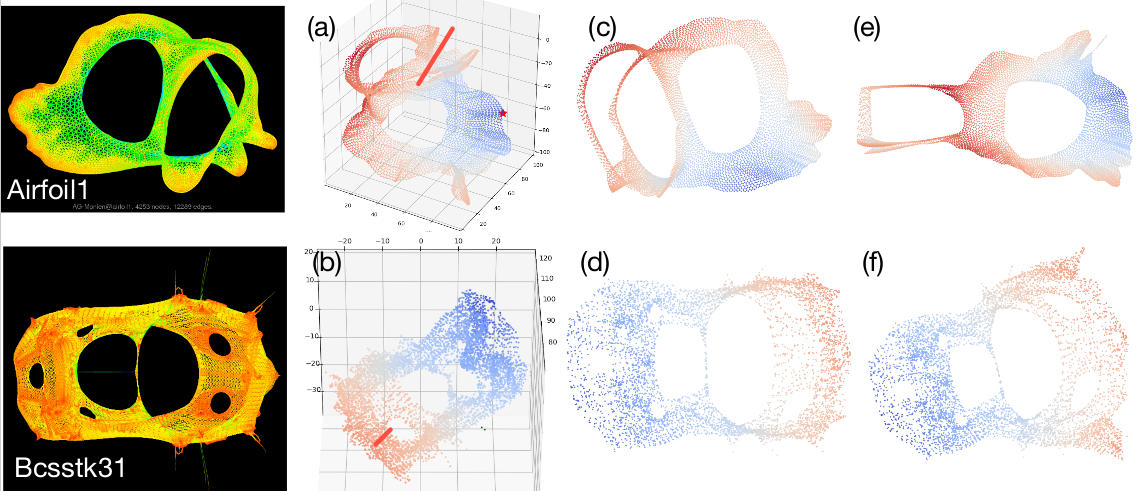}
 \vspace{-2mm}
 \caption{Results without and with manifold tearing for $\airfoil$ (top) and $\bcsstk$ (bottom). (a)-(b) original point clouds with marked cutting location. (c)-(d) Isomap embeddings without tearing. (e)-(f) Isomap embedding with tearing.}
 \label{fig:tearing-matrices}
\end{figure*}

\section{Homology-Preserving Manifold Tearing}
\label{sec:tearing}
Complex data that contain essential loops may or may not be embedded into low-dimensional space without introducing significant distortions. An alternative and complementary approach for homology-preserving DR is through manifold tearing. 
In this section, we give a simple and fast manifold tearing procedure guided by the homology-preserving skeleton of the point cloud data. 
Different from prior work, our procedure tries to cut as few loops as possible, using homology-based quality assessment (described in Section~\ref{sec:evaluation}) while at the same time preserving as much as possible the remaining homological features. 

Suppose we have a point cloud equipped with a pre-computed homology-preserving skeleton (see Section~\ref{sec:landmarking}). 
Our homology preserving tearing process is as follows: 
\begin{itemize} 
\item \emph{Construct neighborhood graph} (same as Isomap). The neighborhood graph is denoted as $G$. A slightly larger $k$ can be chosen to account for the tearing process (optional). 
\item \emph{Tear the neighborhood graph}. A cut plane can be specified based on the homology-preserving skeleton. Specifically, a cut location can be specified on an edge of the skeleton, which then defines a cut plane that is orthogonal to the edge to be cut. An edge that spans a pair of nodes on the opposite side of the cut plane is removed from $G$, resulting a new graph $G'$.
\item \emph{Compute shortest paths}. Compute shortest paths between all nodes in $G'$ and obtain a geodesic distance matrix $D_{G'}$. 
\item \emph{Apply Isomap} to $D_{G'}$. 
\end{itemize}
The above process is exploratory in nature, that is, we can use different evaluation criteria of the resulting embeddings to rank the potential cut locations. In this paper, we use the number of significant homology classes as the criterion. 
In addition, we envision such a process could be embedded into any interactive visualization framework for DR that involves human-in-the-loop. 
\section{Results}
\label{sec:results}

We present examples in this section illustrating that we can achieve homology preservation while at the same time maintaining (and sometimes improving) distance preservation.
We explore complementary procedures with a common goal: 
homology-preserving manifold landmarking and manifold tearing, both aided by the use of a small homological skeleton. 
Such a skeleton is compact, fast to extract and does not introduce significant distance distortions. 

 \begin{figure*}[tb]
 \centering 
 \includegraphics[width=0.8\textwidth]{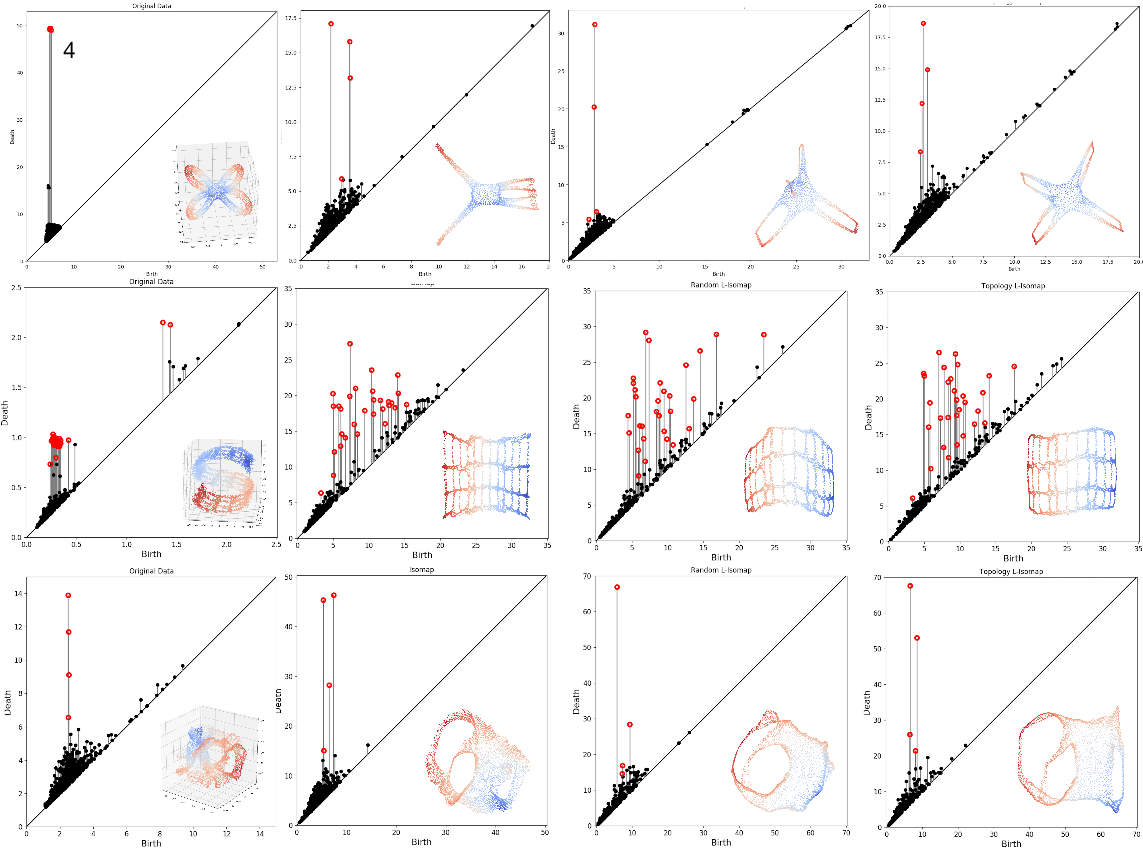}
 \vspace{-4mm}
 \caption{Homology-based quality assessment of DR for $\octa$ (row 1), $\fishing$ (row 2) and $\fourelt$ (row 3). For each each, from left to right: persistence diagrams for the original data,  Isomap embedding, random L-Isomap embedding and homological L-Isomap embedding.}
 \label{fig:pd-all-3}
\end{figure*}

\subsection{Data}
For manifold landmarking, we demonstrate our technique with datasets that contain nontrivial homology. 
Simplify put, datasets with loops are more likely to benefit significantly from our technique. 

\begin{table}
\small
\centering
\label{my-label}
\begin{tabular}{|l|l|l|l|l|l|l|l|l|}
\hline
 &  \emph{octa}  & \emph{fish.}  & \emph{4elt} & \emph{cyl.-3} & \emph{cyl.-5} & \emph{air.} & \emph{bcsstk} & \emph{mice}\\ \hline
$|\Xspace|$ &  2994  & 6188 & 7807 & 2K & 2K & 8034 & 8030 & 674  \\ \hline
$|\Xspace_L|$  & 76 & 63 &  18 & 54 & 82 & 60 & 53 & 18 \\ \hline
\end{tabular}
\vspace{2mm}
\caption{The number of landmarks $|X_L|$ for each dataset of size $|X|$.}
\vspace{-6mm}
\end{table}

$\octa$ is a point cloud sampled from a mesh of octahedron handles. The original mesh contains up to $41K$ vertices. 
$\fishing$ is a synthetic, noisy point cloud sampled from a ``S"-shaped surface that contains $3 \times 11$ irregular holes. 
$\fourelt$ is derived from a 3-dimensional embedding of the 4elt graph used in~\cite{GansnerKorenNorth2005}. The original graph  from~\cite{Walshaw2000} contains $15606$ nodes and $45878$ edges, and is a mesh created to study fluid flow around a 4-element airfoil. As stated in~\cite{GansnerKorenNorth2005}, the original graph ``exhibits extreme variation in the spatial density of nodes". 

Finally, $\mice$ dataset contains $300$-dimensional point clouds derived from time-varying temperature measurements of pregnant mice~\cite{SmarrZuckerKriegsfeld2016}. The point cloud is generated by standard delayed window embedding with a window size of $300$ in signal processing.   
We run our experiment on a particular pregnant mouse that is not jet-lagged, and hope to detect and preserve $1$-dimensional homological features in the input space that capture periodicity in the signal. 

For manifold tearing, we use datasets that contain essential loops for demonstration. 
$\cylinderthree$ is a point cloud sampled from a cylinder with 3 holes carved out, and $\cylinderfive$ is created similarly. 
$\airfoil$ comes from a 2-dimensional finite element problem under the  AG-Monien Matrix group from the SuiteSparse Matrix Collection~\cite{DiekmannPreis}. 
$\bcsstk$ is derived from a 3-dimensional embedding of a stiffness matrix for automobile component~\cite{DuffGrimesLewis1989}.

\subsection{Dimensionality Reduction with Manifold Landmarking}

\paragraph{Results and evaluation with persistence diagrams.}
For each dataset, 2-dimensional embeddings obtained using homological L-Isomap are compared with Isomap and random L-Isomap in Fig.~\ref{fig:teaser}. 

Evaluation using persistent Betti numbers are illustrated by the $1$-dimensional persistence diagrams in Fig.~\ref{fig:pd-all-3}. 
We determine the number of persistent (significant) features by looking at the separation between points in the diagram. 
Suppose each dataset contains $m$ persistent features in the original point cloud, then top $m$ features with the highest persistence are marked in red within the persistence diagram associated with each embedding.

\begin{figure}[!t]
 \centering 
 \includegraphics[width=0.5\columnwidth]{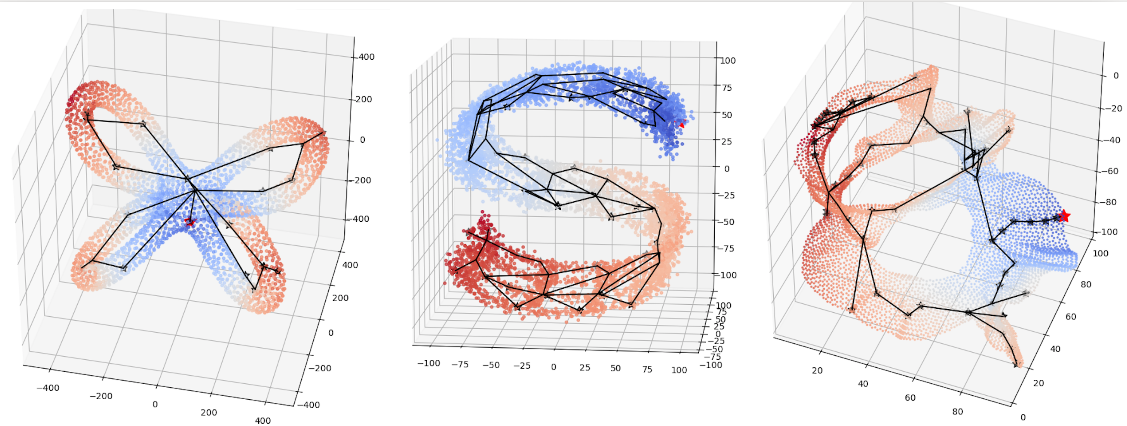}
 \vspace{-2mm}
 \caption{Homological skeletons for $\octa$ (left) using $21$ landmarks, $\fishing$ (middle) and $\fourelt$(right).}
 \label{fig:skeleton-landmarking}
\end{figure}

For $\octa$, as shown in Fig.~\ref{fig:pd-all-3} (row 1), the original data contains 8 significant features, 4 of which (colored red) correspond to the visible loops via embeddings (4 other features are the interior tunnels within each handle).  
All 4 of the red features are preserved (i.e. they remain significant, that is, well-separated from the diagonal of the persistence diagram) using homological L-Isomap. Isomap preserves 3 loops while random L-Isomap preserves only 2. 
For more detailed analysis, it is remarkable to see that using only $21$ landmarks, the homological skeleton is able to summarize the homological features reasonably well (Fig.~\ref{fig:skeleton-landmarking}, left). 

For $\fishing$, as shown in Fig.~\ref{fig:pd-all-3} (row 2), the original data has 33 significant features in the persistence diagram; and both Isomap and homological L-Isomap perform comparatively in terms of preserving the shape of each feature in the embeddings. However, homological L-Isomap uses only $66$ points as landmarks (roughly $1\%$ of the size of the point cloud), and is therefore more computationally efficient. Furthermore, its homological skeleton in Fig.~\ref{fig:skeleton-landmarking} (middle) captures the underlying homological feature pretty well.  

For $\fourelt$, as shown in Fig.~\ref{fig:pd-all-3} (row 3), the original data contains 4 significant features; 3 of which are readily visible in the 3-dimensional embedding of its homological skeleton in Fig.~\ref{fig:skeleton-landmarking} (right). Both Isomap and homological L-Isomap preserve these features reasonably well, while random L-Isomap preserves only 2. In addition, homological L-Isomap does slightly better in preserving the shape of a couple of features. 

\begin{figure}[h]
 \centering 
 \includegraphics[width=0.5\columnwidth]{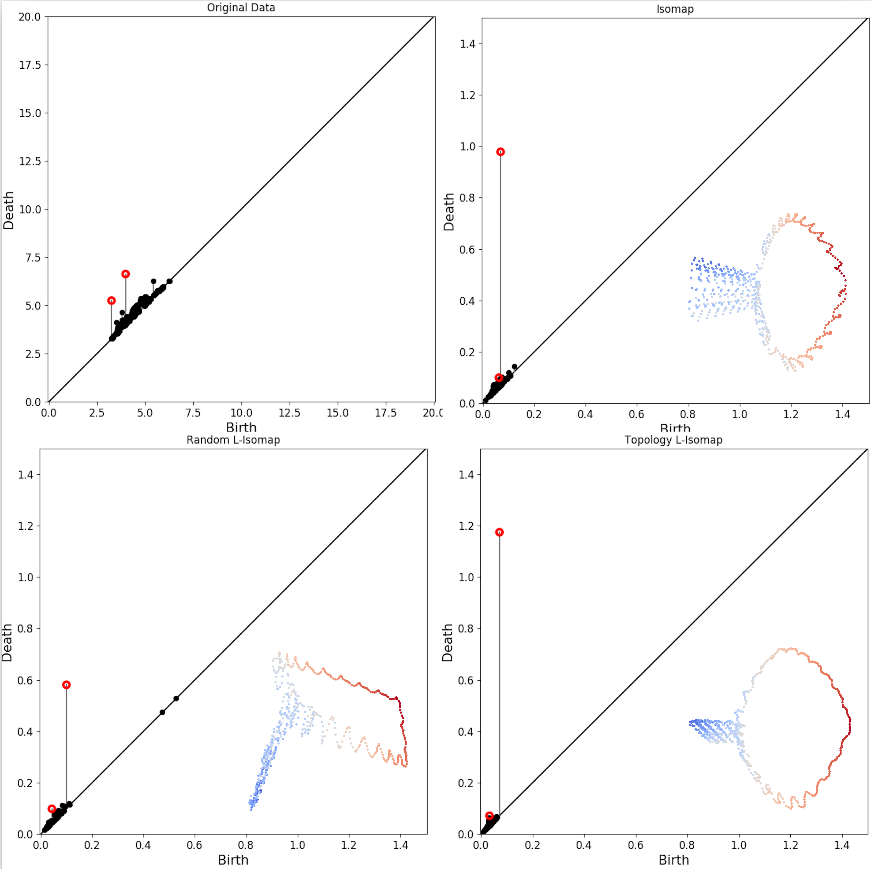}
 \caption{For $\mice$, persistence diagrams for original data (a), Isomap embedding (b), random L-Isomap embedding (c) and homological L-Isomap embedding (d) are combined with DR results.}
 \label{fig:landmarking-mice-pd}
\end{figure}

For $\mice$, we combine the results of DR with $1$-dimensional persistence diagrams in Fig.~\ref{fig:landmarking-mice-pd}. 
There are 2 significant features in the original 300-dimensional input space. Such features likely correspond to periodicity in the temperature profile of the mice that corresponds to circadian or ultradian rhythms. 
Both Isomap and homological L-Isomap perform comparatively in terms of preserving the most dominant feature, while homological L-Isomap only uses a small fraction of the points as landmarks. On the other hand, random L-Isomap is able to detect the significant feature but does not preserve its shape as well.

\paragraph{Quality assessment with residual variance.} 
We also assess the quality of DR using quality  measures introduced in Section~\ref{sec:evaluation}, in particular, the $\RV$ measure. 
$\WD_0$ and $\WD_1$ measures are also computed (but not reported here), as they are partially encoded by persistence diagrams in the previous section.

\begin{figure}[tb]
 \centering 
 \includegraphics[width=0.5\columnwidth]{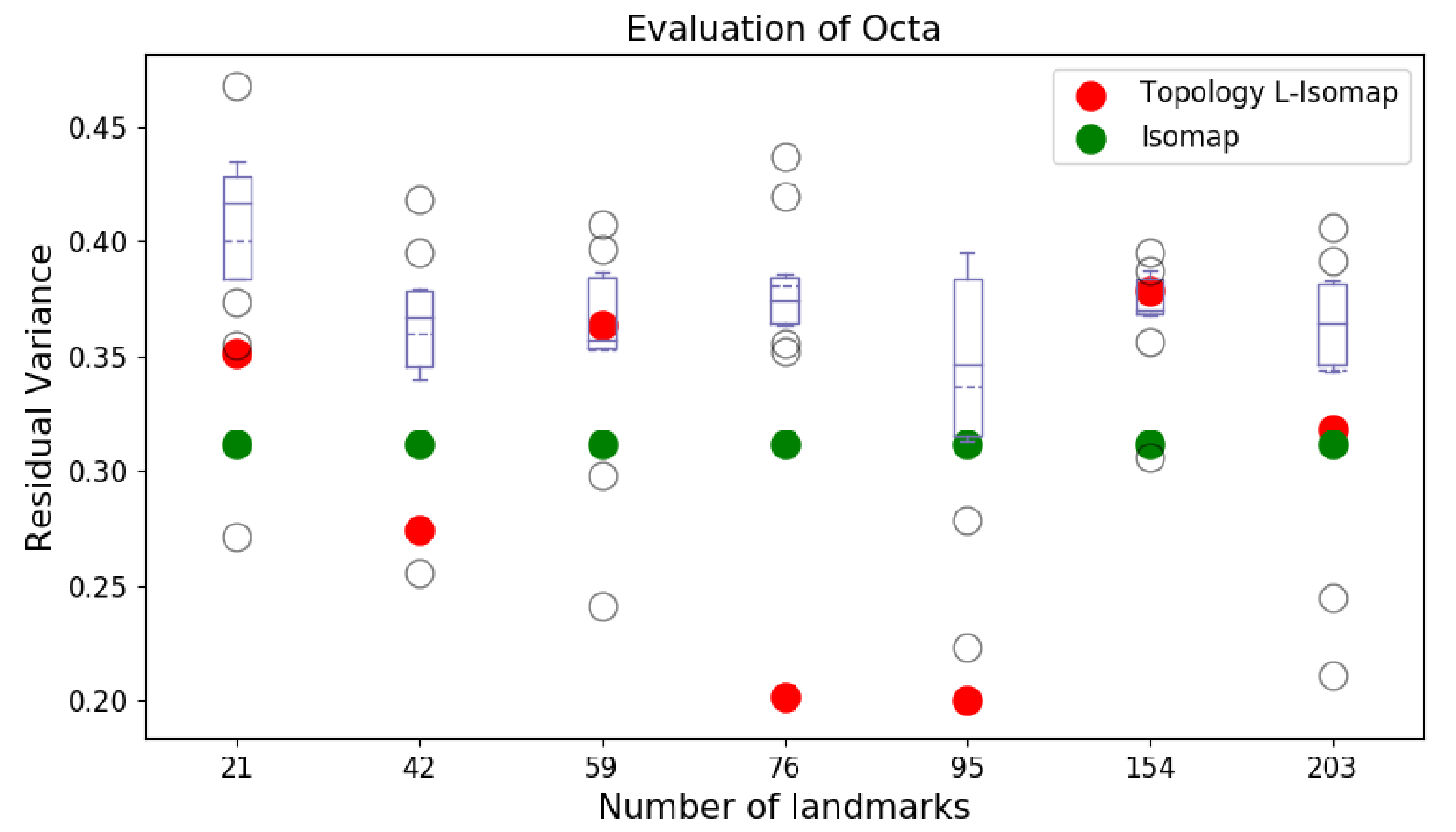}
 \vspace{-2mm}
 \caption{Quality assessment of embeddings using residual variance for $\octa$.}
 \label{fig:rv-octa}
\end{figure}

For $\octa$, we evaluate the quality of each embedding using the $\RV$ measure by varying the number of landmarks, see Fig.~\ref{fig:rv-octa}.
As the number of chosen landmarks increases, we are interested in how well homological L-Isomap preserves distances, when compared with Isomap and random L-Isomap. 
For a fixed landmark size, the blue box plot corresponds to the $\RV$ measures for 20 instances of random L-Isomap, each drawing landmarks randomly from a fixed point cloud: solid blue line in the box plot is the median, dotted blue line is the mean, the boundary of the box is the standard deviation, and black hollow circles are outliers.
Solid red circles are RV measures for homological L-Isomap (denoted as Topology L-Isomap in the figure), while solid green circles are for Isomap. 

A surprising observation is that homological L-Isomap outperforms Isomap and random 
L-isomap in terms of distance preservation, when the number of landmarks is small (below $100$). 
In fact, homological L-Isomap beats random L-Isomap with just $21$ landmarks and it outperforms Isomap with $42$ landmarks. 
The optimal landmark size that achieves both computational efficiency and quality is at around $76$ landmarks. 
When the number of landmarks goes beyond $150$, homological L-Isomap does not seem to have an obvious advantage over other methods. 
In fact, at $203$ landmarks, homological L-Isomap performs comparably with Isomap. 
This is not surprising, with a large number of landmarks, both L-Isomap and Isomap preserve the geometry of the data equally well. 

\begin{figure}[tb]
 \centering 
 \includegraphics[width=0.5\columnwidth]{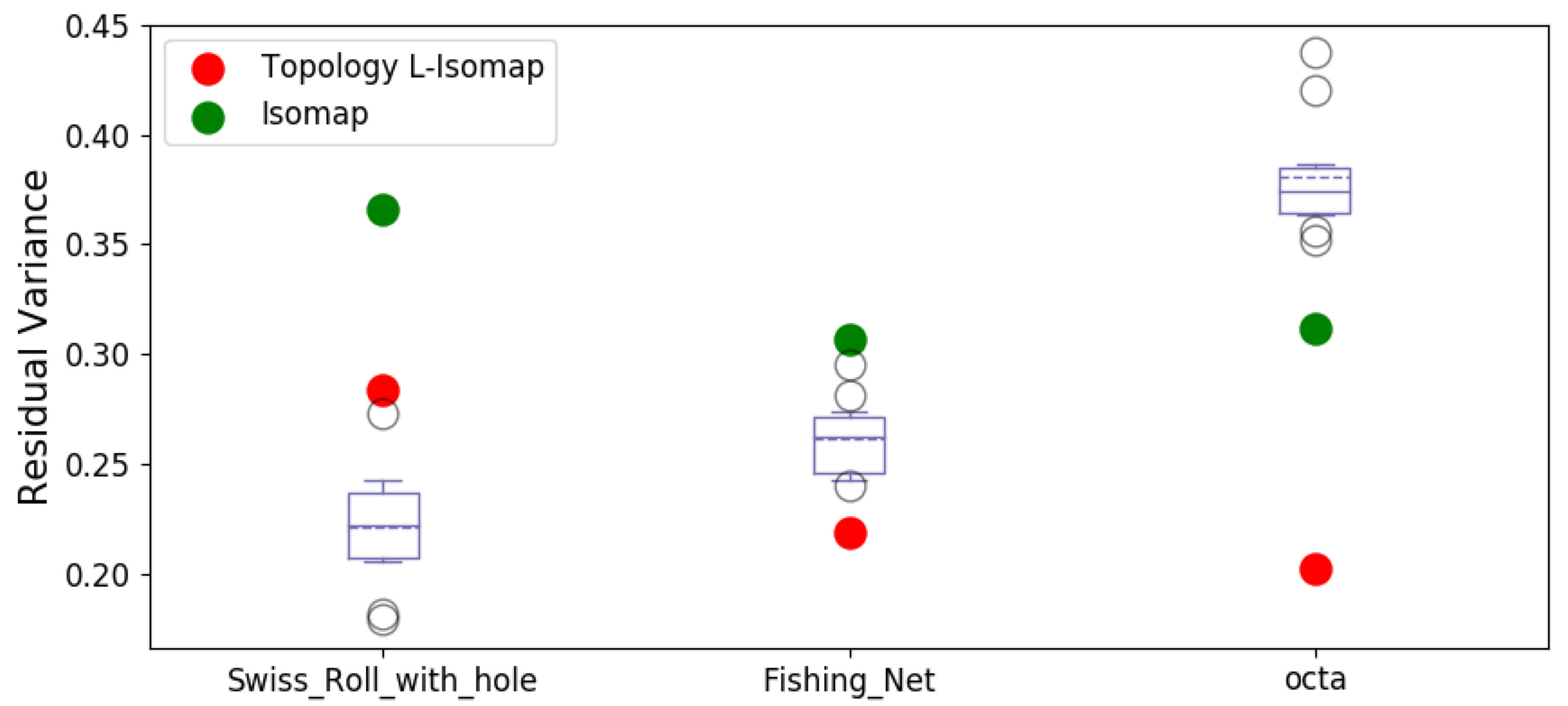}
 \vspace{-2mm}
 \caption{Quality assessment of embeddings using residual variance for different datasets, including $\swissroll$, 
$\fishing$ and $\octa$.}
 \label{fig:rv-swiss-octa-fishing}
\end{figure}

We also compare several datasets, the \emph{swiss roll with a hole}, 
$\fishing$ and $\octa$, using their respective optimal landmark size, in FIg.~\ref{fig:rv-swiss-octa-fishing}. 
Notice that homological L-Isomap outperforms the others for both $\fishing$ and $\octa$, while it does not do well with $\swissroll$.
Intuitively, homological L-Isomap performs best when the data is complex, and has nontrivial homological features. 
In this case, both $\fishing$ and $\octa$ are a lot more complex and homologically interesting than the $\swissroll$.
 
\subsection{Dimensionality Reduction with Manifold Tearing}
\begin{figure}[h]
 \centering 
 \includegraphics[width=0.5\columnwidth]{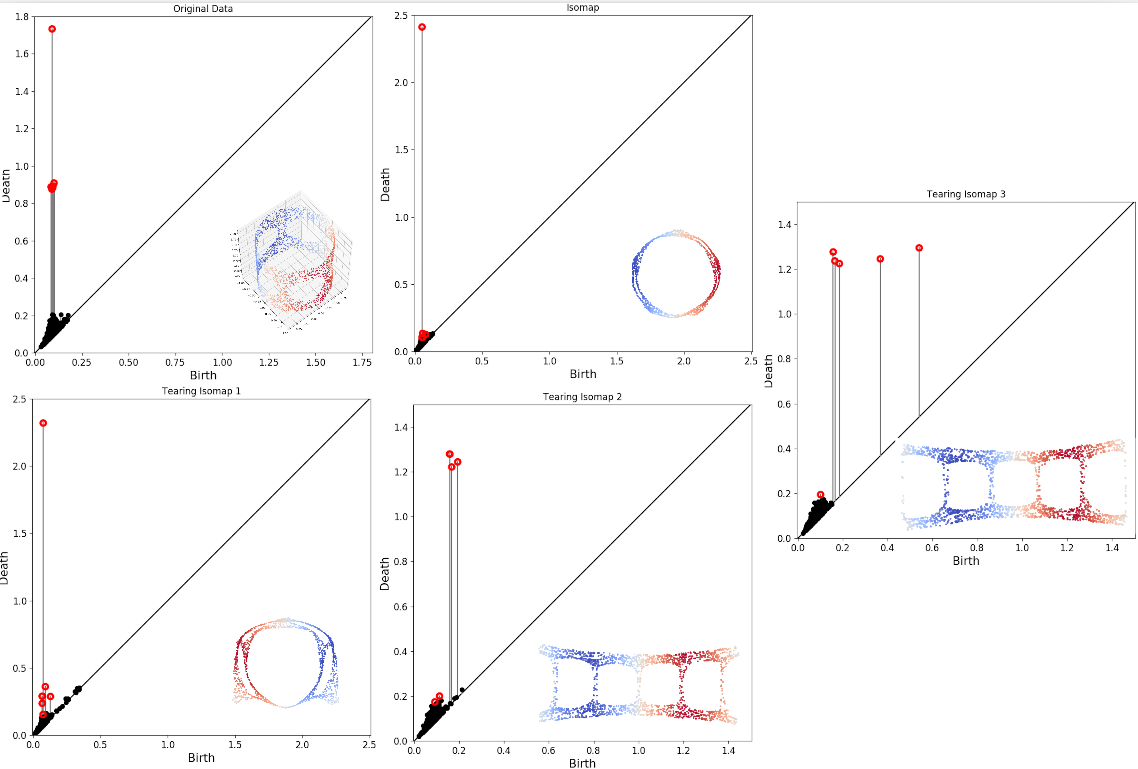}
 \vspace{-4mm}
 \caption{Homology based quality assessment of DR for $\cylinderfive$.}
 \label{fig:pd-cylinder}
\end{figure}

\begin{figure}[h]
 \centering 
 \includegraphics[width=0.5\columnwidth]{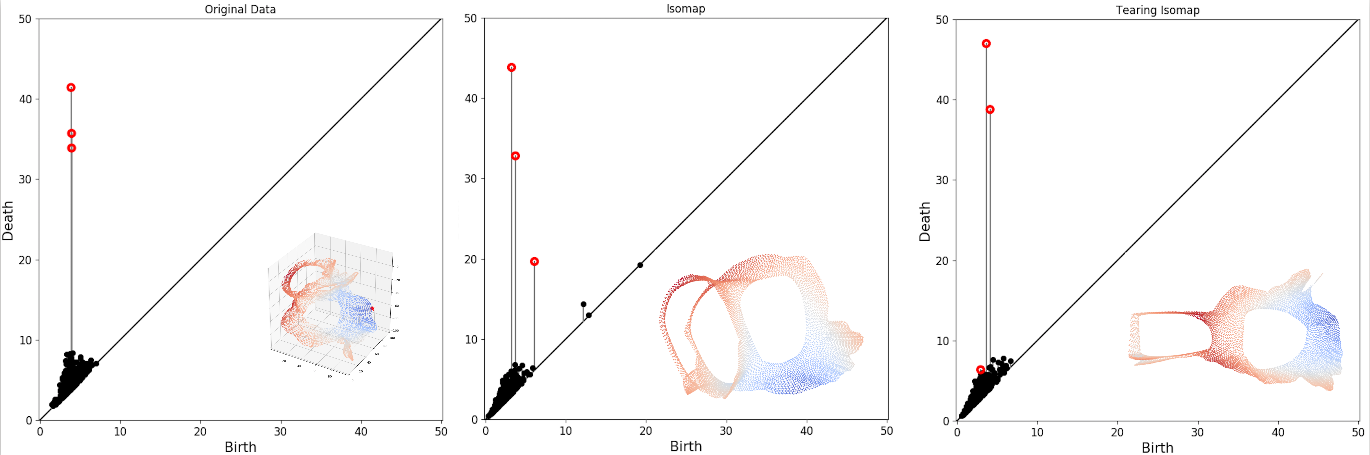}
 \vspace{-6mm}
 \caption{Homology based quality assessment of DR for $\airfoil$.}
 \label{fig:pd-airfoil}
\end{figure}

\begin{figure}[tb]
 \centering 
 \includegraphics[width=0.5\columnwidth]{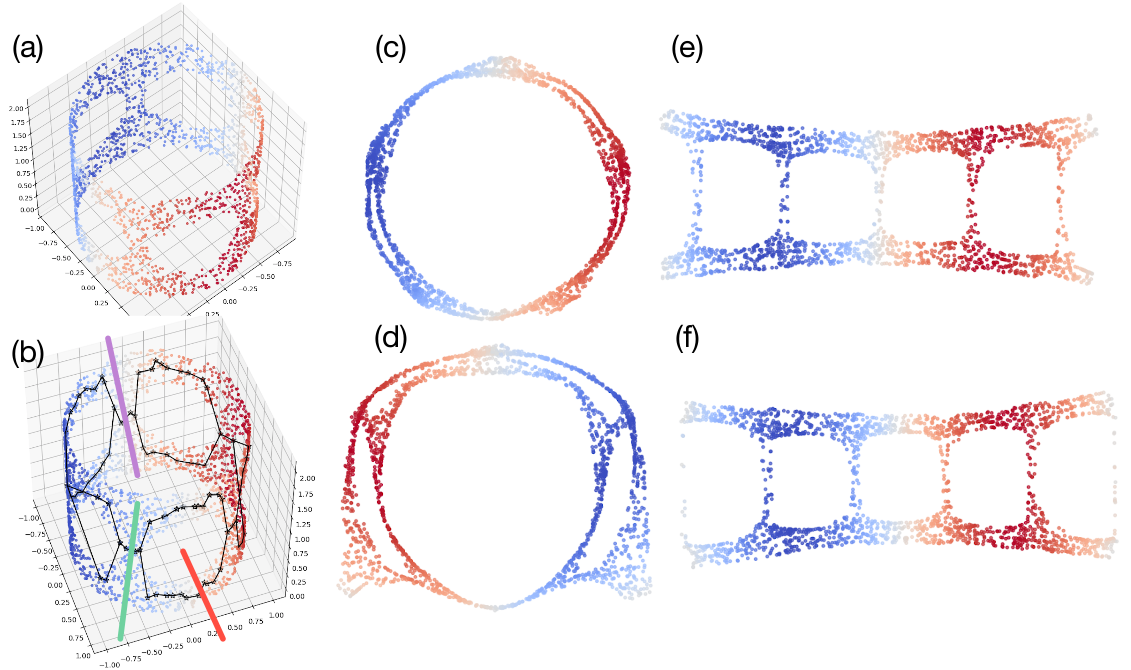}
 \vspace{-2mm}
 \caption{Results without and with manifold tearing for $\cylinderfive$.
 (a) Original point cloud. (b) Homological skeleton with 3 cutting options colored red, purple and green. (c) Isomap embedding without tearing. (d) Partial tearing with the red option. (e) Non-optimal tearing with purple option. (f) Optimal tearing with the green option.}
 \label{fig:tearing-cylinder-5}
\end{figure}

Manifold tearing results are shown in Fig.~\ref{fig:tearing-cylinder-5} for $\cylinderfive$ and Fig.~\ref{fig:tearing-matrices} for $\airfoil$ and $\bcsstk$ respectively. 
See Fig.~\ref{fig:pd-cylinder} and Fig.~\ref{fig:pd-airfoil} for quality assessment using persistence diagrams.  

Given a homological skeleton for $\cylinderfive$, we apply multiple cutting options to the skeleton and rank the resulting embeddings by homology preservation. As shown in Fig.~\ref{fig:tearing-cylinder-5}, without manifold tearing, the Isomap embedding destroys 5 out of 6 homological features, while optimal tearing preserves 5 out of 6 persistent homological features. 

While Isomap preserves reasonably well the 3 persistent features for $\airfoil$, manifold tearing further preserves 2 of the 3 homological features if we are willing to destroy one of them (Fig.~\ref{fig:tearing-matrices} top, and Fig.~\ref{fig:pd-airfoil}). 

\begin{figure}[tb]
 \centering 
 \includegraphics[width=0.5\columnwidth]{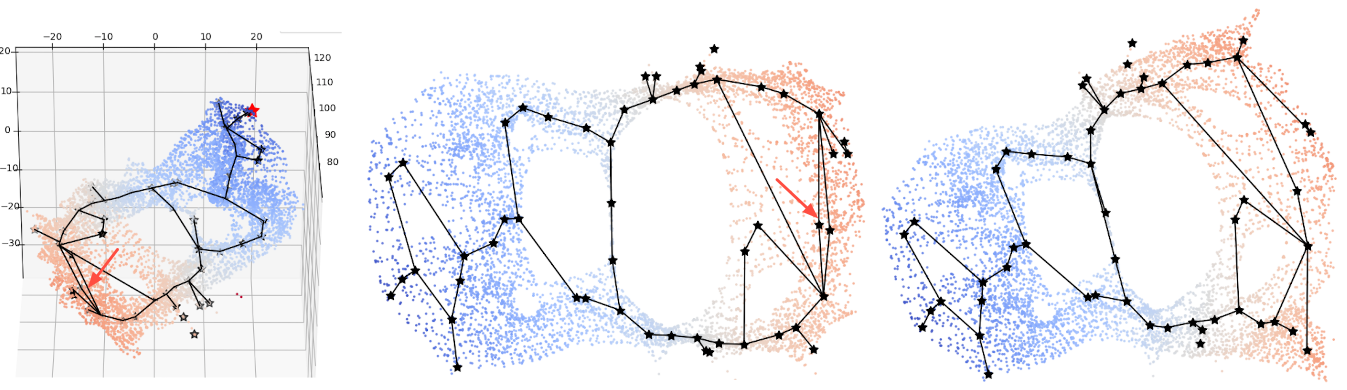}
 \vspace{-2mm}
 \caption{Using homological skeleton to aid manifold tearing for 
 $\bcsstk$. The location marked by the red arrow is where skeleton cutting takes place.}
 \label{fig:tearing-bcsstk31-skeleton}
\end{figure}

For the case of $\bcsstk$, we can focus on manifold tearing by cutting a short edge in the homological skeleton of $\bcsstk$ as shown in~Fig.\ref{fig:tearing-bcsstk31-skeleton}, therefore ``open up" the space further to reveal more geometric structures of the data.  

\section{Discussion}
\label{sec:discussion}
We demonstrate in this paper that we can achieve homology preservation while maintaining and possibly improving distance preservation using homological L-Isomap and (almost) homology-preserving manifold tearing. 
Many research questions remain. In particular, we are interested in exploring higher-dimensional homological skeletons~\cite{VerovssekKurlinLesnik2017} for DR to preserve homological features beyond $1$-dimensions.

\end{document}